\relax
\documentstyle [art12]{article}
\begin {document}
\bibliographystyle {plain}
%\tableofcontents

\title{\bf Superconductivity in a Spin Liquid - a One Dimensional Example}
\author {D.G. Shelton and A. M. Tsvelik}
\maketitle
\begin {verse}
$Department~ of~ Physics,~ University~ of~ Oxford,~ 1~ Keble~ Road,$
\\
$Oxford,~OX1~ 3NP,~ UK$\\
\end{verse}
\begin{abstract}
\par
We study a one-dimensional model of interacting
conduction electrons with a two-fold degenerate  band away
from half filling.
The interaction includes an   on-site Coulomb repulsion and
Hund's rule coupling.
We show that such
one-dimensional
system has a divergent Cooper pair susceptibility at T = 0,
provided the Coulomb interaction $U$ between
electrons on  the same orbital and the modulus of the
Hund's exchange integral $|J|$ are
larger than the interorbital Coulomb interaction. It is remarkable
that the superconductivity can be achieved for {\it any} sign of $J$.
The opening of spectral gaps makes  this state stable
with respect to direct electron
hopping between the orbitals.
The scaling dimension of the superconducting order
parameter is found  to be between  1/4 (small $U$) and
1/2 (large $U$). Possible experimental realizations of such systems
are discussed.
\end{abstract}

%\newpage
\sloppy
\par
\section{Introduction}

 The idea that a doped spin liquid state favours superconductivity
(SC) belongs to Anderson \cite{phil}. This idea was introduced  in the
context of quasi-two-dimensional copper oxide
superconductors which are believed to be adequately described by the
t-J model.  However, it has proved very difficult to reliably analyse
this model on a square lattice. Recently it has been suggested\cite{zhang},
\cite{rice}  that SC might  appear in one dimensional doped spin
liquids which are much easier to analyse\cite{1}.
In this letter we further develop these ideas and give a
more detailed description of the SC spin liquid state
suggested in these papers.

 We study the model of two Hubbard chains with the Hund's
coupling  between the chains:
\begin{eqnarray}
H &=& H_{Hub} + \sum_r\left(\frac{u}{2}\rho_{1,r}\rho_{2,r}
+ 2J{\bf S_{1,r}}{\bf S_{2,r}}\right)\label{vint}\\
H_{Hub} &=&- \frac{D}{2}\sum_{r,a,\alpha}
(c^+_{a,\alpha, r + 1} c_{a,\alpha, r}
+ H. c.) + U\sum_{r,a}c^+_{r,a,\uparrow}
c_{r,a,\uparrow}c^+_{r,a,\downarrow}c_{r,a,\downarrow}
\end{eqnarray}
where $c^+_{a,\alpha,r}, \: c_{a,\alpha, r}$ are electron creation and
annihilation operators on the site $r$,
$\alpha = \pm 1/2$ and $a = 1,2$ are electron spin and  chain
(orbital)
indices respectively, and
\begin{eqnarray}
\rho_a = \sum_{\alpha}c^+_{a,\alpha}c_{a,\alpha}; \: {\bf S} =
\frac{1}{2}\sum_{\alpha,\beta}c^+_{a,\alpha}{\bf
\sigma_{\alpha\beta}}c_{a,\beta}\label{dens}
\end{eqnarray}
where $\sigma^a$ are the Pauli matrices. Electronic mechanisms give
ferromagnetic exchange (Hund's rule); an antifferromagnetic $J$
can be generated by  the Jahn-Teller effect (see the discussion
later).
Later we shall present the
arguments that the model (1) may describe quasi-one-dimensional
materials based
on chains of C$_{60}$ molecules.

 The model (\ref{vint}) does not include a direct hopping between the
orbitals because such terms
are irrelevant when there are spectral gaps.  We shall
demonstrate that at $U >> u$
such gaps exist for all excitations except
for   the symmetric
charge density mode. Since a single electron hopping
process inevitably involves modes with spectral gaps,
the direct hopping becomes energetically costly and therefore
irrelevant. It turns out that if the  certain conditions on $U, u$ and
$J$ are met, the
ground state of the model (1,2)  has strongly
enhanced SC fluctuations. Due to the presence of spectral
gaps in a single pair of chains (further we shall sometimes
call such pair a ``ladder''),
the only excitations which can tunnel between the ladders are Cooper
pairs. As we shall demonstrate later, the scaling dimension of the SC
order parameter is small which makes the pair susceptibility more
singular than in BCS superconductors and thus  increases the
critical temperature of a  three dimensional system made of weakly
interacting ladders.

  Khveshchenko and Rice\cite{rice} in their analysis of the model
(1,2) have used  bosonization in the weak coupling limit
combined with  the renormalization group approach (RG)
(later this model has been also studied by Fujimoto and
Kawakami\cite{kawakami} with the same qualitative results). We suggest
an alternative
approach which allows us to proceed directly to the strong
coupling fixed point
avoiding  a cumbersome RG analysis of the weak coupling limit.
The
suggested reformulation of the Hamiltonian  enables us to expand
around
the strong coupling point. The reason this
representation  is so useful
is that it elucidates the inherent symmetries of the Hamiltonian.
We demonstrate that in order to realize  the mechanism of SC  suggested
in \cite{zhang},\cite{rice}, one needs  to have a
hierarchy of couplings where
the on-chain repulsion must be  the strongest
one. Together with the interchain exchange interaction it
creates  spectral gaps for all excitations except for  the symmetric
charge mode. The interchain Coulomb repulsion $u$ has a potentially
damaging effect on SC establishing a lower  limit on
the magnitude of $|J|$. It is intriguing
that the conditions for SC can
be met for both signs of $J$.
The SC state is stable with respect to direct hopping if the
hopping amplitude $t$ is much smaller than a sum  of the gaps.

\section{Bosonization and Introduction of Majorana Fermions.}

 As usual, interactions between  electrons in one-dimension cannot be
treated perturbatively. Hence we have to  resort to non-perturbative
methods, the most popular  among which  is bosonization.
Since the Hubbard model is exactly
solvable for any $U/D$ and doping, the bosonization approach
requires only  a smallness of the Hund's rule couplings  $u$ and  $J$. The
behaviour of the model (\ref{vint}) at half filling has been
already well studied \cite{ners}. In this paper we assume a fair
amount
of doping.

 We begin by bosonizing the charge and the spin density operators
(\ref{dens}). According to Frahm and Korepin(\cite{frahm})
the continuous limit of these
operators is given by (here we omit the orbital  indices)
\begin{eqnarray}
\rho(x) = J_R + J_L +
%% FOLLOWING LINE CANNOT BE BROKEN BEFORE 80 CHAR
2\cos(\sqrt{2\pi}\Phi_s)\left[a_{c}e^{2\mbox{i}k_Fx}\exp(\mbox{i}\sqrt{2\pi}\Phi_c)
+ H. c.\right]\nonumber\\
 + \left[a_{u}e^{4\mbox{i}k_Fx}\exp(\mbox{i}\sqrt{8\pi}\Phi_c) + H.
c.\right]\label{rho}\\
{\bf S}(x) = {\bf J}_R + {\bf J}_L +
\left[a_{s}e^{2\mbox{i}k_Fx}\vec \sigma(x) + H. c.\right]\label{S}\\
\sigma^3 = \mbox{i}\sin(\sqrt{2\pi}\Phi_s)e^{-
\mbox{i}\sqrt{2\pi}\Phi_c}; \: \sigma^{\pm} =
\exp[\mbox{i}\sqrt{2\pi}(- \Phi_c \pm \Theta_s)]\label{sigma}
\end{eqnarray}
where the corresponding bosonic fields
are governed by the following Hamiltonian:
\begin{eqnarray}
H_0 &=&
\frac{1}{2}\int \mbox{d}x\left[v_cK_c(\partial_x\Theta_c)^2 +
\frac{v_c}{K_c}(\partial_x\Phi_c)^2 + v_s(\partial_x\Theta_s)^2 +
v_s(\partial_x\Phi_s)^2\right]\label{ho}
\end{eqnarray}
where the operators satisfy the standard commutation relations:
\begin{eqnarray}
[\Phi_a(x),\Theta_b(y)] = \mbox{i}\theta(x - y)\delta_{ab}
\end{eqnarray}
and $a_c, \: a_u$ and $a_s$ are numbers which numerical values
are known only in some limiting
cases (we shall return to this issue later).
In particular, $a_u = 0$ for a free
electron gas. The charge and spin currents
$J_{R,L}$ and ${\bf J_{R,L}}$ obey the corresponding Kac-Moody
algebras (see, for example, \cite{drouffe}). The constant $K_c$
and charge and spin velocities $v_c, \: v_s$ depend on
the interaction and doping such that $1/2 < K_c <
1$. The lowest limit is achieved for $U = \infty$ and $K_c = 1$
corresponds to the non-interacting case. The fact that the
corresponding spin constant remains unrenormalized ($K_s = 1$)
reflects the SU(2) symmetry of the individual Hubbard chain (\cite{frahm}).

 Substituting expressions (\ref{rho}, \ref{S}, \ref{sigma}) into
Eq.(\ref{vint}) and keeping only non-oscillatory terms, we get the
following expression for the interaction density:
\begin{eqnarray}
V_{int} &=& V + {u\over 2}[J_{R1}+J_{L1}][J_{R2}+J_{L2}]
+2J[{\bf J}_{R1}+{\bf J}_{L1}][{\bf J}_{R2}+{\bf J}_{L2}]\\
V &=&
2\cos(\sqrt{4\pi}\Phi_c^-)[u|a_c|^2(\cos\sqrt{4\pi}\Phi_s^+ +
\cos\sqrt{4\pi}\Phi_s^-) \nonumber\\
&+& J|a_s|^2(\cos\sqrt{4\pi}\Phi_s^+ -
\cos\sqrt{4\pi}\Phi_s^- + 2\cos\sqrt{4\pi}\Theta_s^-)] \label{expon}
\end{eqnarray}
The $4k_F$-scattering  contributes the  irrelevant operator
$\cos{4\sqrt\pi\Phi_c^-}$ which we omit. Here we have introduced
symmetric and antisymmetric combinations of the bosonic fields:$
\Phi_{c,s}^{\pm} = (\Phi^{1}_{c,s}
\pm \Phi^{2}_{c,s})/\sqrt 2$
This transformation leaves the bosonic Hamiltonian (\ref{ho})
invariant.

 The interaction of charge currents can be easily incorporated into
the free Hamiltonian (\ref{ho}) by the change of $K_c$'s and velocities for the
symmetric and antisymmetric charge modes:
\begin{eqnarray}
 K^{\pm}_{c} &=& \left[1 \pm ({u\over 2\pi
v_{c}})\right]^{-1/2} K_{c} \label{kc}
\end{eqnarray}
We shall assume that $K_c^- < 1$, i.e. the interaction in the
antisymmetric charge channel remains repulsive. Then the
scaling dimension of the operator $V$ $d_V = 1 + K_c^- < 2$ is always
smaller than the dimension of the product  of spin currents,
which is a marginal operator. Therefore we omit the latter  term.
In the subsequent
analysis we will take advantage of the fact that all
bosonic exponents in the square brackets in Eq.(\ref{expon}) have the
scaling dimension 1 which means that they  can be expressed as
fermionic bilinears. The corresponding expressions in terms of Majorana (real)
fermions have been derived in Ref.\cite{ners}:
\begin{eqnarray}
(2 \pi a_0)^{-1}\cos\sqrt{4\pi}\Theta_s^- = \mbox{i}(R_1L_1 - R_0L_0), \:
(2 \pi a_0)^{-1}\cos\sqrt{4\pi}\Phi_s^- = \mbox{i}(R_1L_1 +
R_0L_0),\nonumber\\
(2 \pi a_0)^{-1}\cos\sqrt{4\pi}\Phi_s^+ = \mbox{i}(R_2L_2 + R_3L_3)
\end{eqnarray}
where $a_0$ is the short distance cut-off and the fermions satisfy the
following anticommutation relations:
\begin{eqnarray}
\{R_a(x),R_b(y)\} = \frac{1}{2}\delta(x - y)\delta_{ab}; \:
\{L_a(x),L_b(y)\} = \frac{1}{2}\delta(x - y)\delta_{ab},\:
\{R_a(x),L_b(y)\} = 0
\end{eqnarray}
Being recast in new terms
the Hamiltonian (\ref{ho}) becomes
\begin{eqnarray}
H_0 = \frac{1}{2}\sum_{\pm}\int
\mbox{d}x\left[v^{\pm}_cK_c^{\pm}(\partial_x\Theta_c^{\pm})^2 +
\frac{v^{\pm}_c}{K_c^{\pm}}(\partial_x\Phi_c^{\pm})^2\right]\nonumber\\
+ \mbox{i}v_s\sum_{a = 0}^3\int \mbox{d}x\left(L_{a}\partial_xL_a -
R_{a}\partial_xR_{a}\right)
\end{eqnarray}
The interaction density acquires the following form:
\begin{eqnarray}
V &=& \mbox{i}\cos\sqrt{4\pi}\Phi_c^-\left[m_1(1 - \delta_{a,0})R_aL_a
- m_2R_0L_0\right]\nonumber\\
m_1 &=& 4\pi a_0(u|a_c|^2 + J|a_s|^2), \: m_2 = 4\pi a_0(3J|a_s|^2 - u|a_c|^2)
\label{v}
\end{eqnarray}
 If $K_c^- < 1$ (that is $u/2\pi v^-_c < 1 - K_c^2$,  which
requires  the  Coulomb
repulsion $U$ to be stronger than $u$!) the interaction (\ref{v})
is relevant making  the
Majorana fermions  and the field
$\Phi_c^-$ massive. The modes $a \neq 0$ and $a = 0$ acquire different masses
which we call $M_t$ and $M_0$ respectively  with $t$ standing for
``triplet''. The triple degeneracy of three fermion modes reflects the
fact that they realize the spin 1 representation of the SU(2) group
\cite{ners}. The mass of the bosonic field $\Phi_c^-$ is denoted
$\tilde M$. The dimensional analysis gives $M_t, M_0, \tilde M \sim J^{1/(1 -
K_c^-)}$.
One can safely assume that the following averages do not
vanish:
\begin{eqnarray}
\langle\cos\sqrt{4\pi}\Phi_c^-\rangle \neq 0; \: \langle R_{a}L_{a}\rangle \neq
0 \label{av}
\end{eqnarray}
In the limit $K_c^- << 1$ this can be rigorously proven by the
perturbation expansion in $K_c^-$. Namely, one should rescale the
field $\Phi_c^-$: $\Phi_c^- = \sqrt{K_c}\tilde\Phi_c^-$ and expand the
cosine term in Eq.(\ref{v}):$\cos\sqrt{4\pi K_c}\tilde\Phi_c^- = 1 - 2\pi
K_c(\tilde\Phi_c^-)^2 + ..$. Then in the first approximation in $K_c$
one obtains
the quadratic
effective Hamiltonian (\ref{heff}). One can check that corrections
coming from higher terms are not singular and contain powers of the small
parameter $K_c$.
The appearance of the averages (\ref{av})
brakes the discrete Z$_2$ symmetry
which is not forbidden in (1 + 1)-dimensions. The corresponding
order parameter is non-local in terms of the original fermionic
fields\cite{ners}.

\section{Superconducting fluctuations.}

 As follows from the previous discussion, the Hamiltonian
for the field $\Phi_c^+$ remains
unaffected such that this field is still a free bosonic field.
The other fields become massive and their
 strong coupling limit can be qualitatively described by the
effective Hamiltonian:
\begin{eqnarray}
H_{eff} = \mbox{i}\int\mbox{d}x v_s\sum_{a = 0}^3\left[ -
R_{a}\partial_xR_{a} +
L_{a}\partial_xL_{a} + (M_t(1 -
\delta_{a,0}) - M_0\delta_{a,0})R_aL_{a}\right]\nonumber\\
+ \frac{1}{2}\int\mbox{d}x[v_c^-(\partial_x\Theta^{-}_c)^2 +
v_c^-(\partial_x\Phi^{-}_c)^2 + \tilde M(\Phi^{-}_c)^2]\label{heff}
\end{eqnarray}
As we have mentioned above this  Hamiltonian becomes exact in the
limit $K_c^- \rightarrow 0$.

 We have deliberately put a minus sign at $M_0$ in Eq.(\ref{heff}).
We shall see that the
correlation length of the SC order parameter is infinite
only provided $M_t$ and $M_0$ in Eq.(\ref{heff})
have the same sign. As follows from
Eq.(\ref{v}), the masses are
\begin{eqnarray}
M_t \sim (u|a_c|^2 + J|a_s|^2)\langle\cos\sqrt{2\pi}\Phi_c^-\rangle; \:
M_0 \sim (3J|a_s|^2 - u|a_c|^2)\langle\cos\sqrt{2\pi}\Phi_c^-\rangle
\end{eqnarray}
Combining this criterion with $K_c^- < 1$ we obtain  the following
criterion for the existence of an infinite correlation
length for SC fluctuations in the model (1,2):
\begin{equation}
K_c^- < 1; \: (u|a_c|^2/|a_s|^2 + J)(3J - u|a_c|^2/|a_s|^2) > 0 \label{criter}
\end{equation}
The coefficients $a_c, a_s$ are known only in the weak coupling limit
$U/D << 1$ where $a_c = a_s$. It is also likely that $|a_s| >> |a_c|$
close to half filling where charge fluctuations are suppressed.
Thus we can resolve  the inequalities (\ref{criter}) explicitly only
in the limit of weak interactions where $K_c \approx 1 - U/4\pi v_c$:
\begin{equation}
U > u; \: J > u/3 \: \mbox{or} \: J < - u
\end{equation}

 Let  $\psi_{R,L}$ be  right- and left moving components of the
original fermions.
Now we shall demonstrate  that if the criterion (\ref{criter}) is
met, the susceptibility of the SC order parameter defined as
\begin{eqnarray}
\Delta = \psi_{R,1,\uparrow}\psi_{L,2,\downarrow} \pm
\psi_{R,2,\uparrow}\psi_{L,1,\downarrow}\nonumber\\
\sim
\exp\mbox{i}\sqrt{\pi}(\Phi_s^+ +
\Theta_c^+)\{\exp[\mbox{i}\sqrt{\pi}(\Theta_s^-+\Phi_c^-)]\pm
\exp[- \mbox{i}\sqrt{\pi}(\Theta_s^-+\Phi_c^-)]\} \label{defin}
\end{eqnarray}
is singular.
The choice of sign in the above expression
depends on the sign of $\langle\cos\sqrt{4\pi}\Phi_c^-\rangle$ (broken Z$_2$
symmetry).
The  order parameter includes exponents of the fields $\Phi_s^+$ and
$\Theta_s^-$ with scaling dimensions 1/4. In the Majorana approach
it is most convenient to express these fields in terms
of the order  ($\sigma$) and disorder ($\mu$) parameter
fields of the Ising model using
the fact that the model of massive Majorana fermions describes
a  two dimensional Ising model off the critical point, where
the fermionic mass is related to the
deviation from $T_c$: $M \sim (T - T_c)$. The
corresponding operators are related as follows\cite{miwa}:
\begin{eqnarray}
\exp(\mbox{i}\sqrt\pi\Phi_s^+) \sim \mu_1\mu_2 +
\mbox{i}\sigma_1\sigma_2; \: \exp(- \mbox{i}\sqrt\pi\Theta_s^-) \sim
\sigma_3\mu_0 -
\mbox{i}\sigma_0\mu_3
\end{eqnarray}
Substituting these expressions into Eq.(\ref{defin}) we get
\begin{eqnarray}
\Delta^{\pm} =
\exp(\mbox{i}\sqrt{\pi}\Theta_c^+)\cos\sqrt{\pi}\Phi_c^-(\mu_1\mu_2 +
\mbox{i}\sigma_1\sigma_2)\left\{
\begin{array}{cc}
\sigma_3\mu_0 \\
-\mbox{i}\sigma_0\mu_3
\end{array}
\right.
\approx \lambda :e^{\mbox{i}\sqrt{\pi}\Theta_c^+}: \label{delta}
\end{eqnarray}
where depending on the sign of $\langle\cos\sqrt\pi\Phi_c^-\rangle$
\begin{equation}
\lambda = \langle\cos\sqrt\pi\Phi_c^-(\sigma_1\sigma_2\sigma_3\mu_0)
\rangle \mbox{or}
\langle\cos\sqrt\pi\Phi_c^-(\mu_1\mu_2\mu_3\sigma_0)\rangle
\end{equation}
 In the
expression (\ref{delta}) we have omitted the
term proportional to  $\sin\sqrt\pi\Phi_c^-$ since the sinus
has a zero average.
As is well known, one of the operators $\sigma$ and  $\mu$ has a nonzero
average off the critical point. Which one  does not vanish depends
on the sign of $(T - T_c)$, that on the sign of mass in
the Majorana representation. In order to have $\lambda \neq 0$,
the mass term of
the zeroeth
Majorana mode should have  a sign opposite to the mass term of the
other modes which gives as criterion (\ref{criter}).

 It is easy to see  that if $M_0M_t < 0$ the order parameter describes
the charge density wave:
\begin{equation}
O_{CDW}(x) = \sum_{\alpha}(\psi^+_{R,1\alpha}\psi_{L,1\alpha} \pm
\psi^+_{R,2\alpha}\psi_{L,2\alpha})
\end{equation}
As follows from  Eq.(\ref{delta}), at T = 0 the correlation function
of Cooper pairs has the following asymptotic
behaviour at
distances $>> M$ ($M$ is the smallest gap):
\begin{eqnarray}
\langle \Delta(x,\tau)\Delta^+(0,0)\rangle &\sim& M^2{1\over (M|\tau
+\mbox{i}{x\over v_c}|)^{1/2K_c^+}}
\end{eqnarray}
with $K_c^+$ defined by Eq.(\ref{kc}).  Since  $K_c$
varies from $1$ at $U = 0$ to ${1\over 2}$
at $U =\infty$, the scaling dimension of the order parameter
$1/4 < d = \frac{1}{4K_c^+} < 1/2$ and the Fourier
transformation of the correlation function (pairing susceptibility)
always diverges at $\omega, q = 0$. At finite temperatures there will be
finite number of kinks interpolating between vacua with
$\langle\cos\sqrt{4\pi}\Phi_c\rangle > 0$ and $< 0$. Thus the Z$_2$
symmetry is restored with the finite correlation length $\xi \sim
\exp(M_k/T)$, where $M_k \sim M_0/K_c$ is the kink's mass. This
exponentially large correlation length will be completely overshadowed
by the correlation length of the $\Phi_c^+$-field: $\xi_c \sim 1/T$.

\section{Physical Realization}

 One posible group of experimental systems where the described superconducting
mechanism may work are  materials based on chains of C$_{60}$
molecules. An isolated C$_{60}$ molecule in a crystal has
a rich degeneracy and the
Hund's interaction may be generated by  local vibronic modes
(the Dynamical Jahn-Teller effect; see
Ref.\cite{mary} for a review). In  a
structure where C$_{60}$ molecules are
situated on well separated chains a two-fold  local degeneracy may
survive even in
the presence of an intermolecular hopping.
Then the one-dimensional electronic
band is split into a single degenerate band created by the $s, p_z$ orbitals
and a doubly degenerate band created by $p_x, p_y$ orbitals. The
measurements performed for RbC$_{60}$ give many indications of
a quasi-one-dimensional behaviour \cite{chauvet}. The direct
band structure calculations done for A$_{n}$C$_{60}$-compounds with
A = K, Rb show, however,  that the resulting
electronic bands have a three dimensional character \cite{mele}.
Nevertheless,  even if these calculations are
correct in the given case,
one may expect that it would not be too difficult to produce truly
one-dimensional C$_{60}$-based materials. The authors of Ref.\cite{mele}
remark that the interchain hopping is very
sensitive to the distance between the chains  and therefore
the one-dimensional limit ``might well be achieved by doping with
substantially larger species''. These  authors further suggest
creating  chemical compounds of  C$_{60}$ with
``spacer molecules'' such as NH$_3$.

 The model  describing
a one-dimensional
chain of Jahn-Teller molecules with electrons belonging
to two degenerate local orbitals hopping  along the chain and
interacting with local optical (vibronic) modes of the molecules was
introduced by Manini {\it et al}.\cite{tosatti}.
The Hamiltonian of electron-phonon interaction
is given by
\begin{eqnarray}
H &=& \sum_{r, n = x,y}\left[
gc^+_{r,a,\alpha}\hat{\sigma^n}_{ab}c_{r,b,\alpha}u_n(r)  -
\frac{1}{2M}\frac{\partial^2}{\partial {u_n(r)}^2} +
\frac{ku_n(r)^2}{2}\right] \label{eq:model}
\end{eqnarray}
where $u^x(r), \: u^y(r)$ describe  local vibronic modes. A
single C$_{60}$ molecule has a degeneracy higher than 2,
but it is  assumed  that such high
degeneracy does not  survive in a crystal. Neglecting
all retardation effects related to the kinetic energy of vibrons and
integrating  over the vibronic modes we get the interaction
(\ref{vint}) with $u = J = 2g^2/k$.  As in the standard BCS theory, the
retardation effects
will give us a high energy cut-off of the order of the vibronic
frequency $\omega_0 = (k/M)^{1/2}$. Santoro {\it et al.}\cite{santoro}
have studied this interaction in the approximation
where the Jahn-Teller bound states
were considered first, the hopping was treated as a
perturbation and  the Coulomb interaction was ignored.
Thus  the dynamical nature of the vibronic modes appeared  to be an
essential feature of this approach. As a result they got the scaling
dimention of the SC order parameter equal to 1/4 in
agreement  with our calculations.  The fact
that the two approaches give  the same infrared behaviour is
remarkable, but is by no means unusual. This is a standard feature of
one-dimensional systems where one almost always reproduces strong
coupling results starting from weak coupling. This is due to the fact
that the strong coupling appears eventually in the
renormalization process.

\section{Conclusions}

 A curious property of our results is that SC
requires a strong Coulomb repulsion between electrons on the same
orbital. It is this repulsion which leads to formation of the spin gap
with subsequent freezing out some electronic degrees of freedom. There
is also a charge gap corresponding to antiphase fluctuations of the
orbital charge densities. In the situation where the exchange
interaction is generated by the Jahn-Teller mechanism, this
gap coincides with the spectral gap of
vibrons at $q = 2k_F$.

The Coulomb interaction, however, plays a double role since
the scaling dimension of the order parameter
increases when the Coulomb repulsion grows becoming  equal to 1/2 at
$U = \infty$. The
small scaling dimension of the SC order parameter
increases the  probability of having  a relatively
large transition temperature in
a quasi-three-dimensional system built of weakly interacting
chains. If $t$ is  the interchain Josephson coupling,
then  the dimensional analysis gives for
the temperature of SC transition the estimate
\begin{equation}
T_{\mbox{c}} \sim M_0(t/M_0)^{\eta}; \: \eta = \frac{1}{2 - 2d}
\end{equation}
where $M_0 \sim g^{1 + 1/(1 - K_c^-)}$ is the spin singlet gap.
In the conventional case $d \approx 1$ and the transition temperature
is exponentially small in $M_0/t$, but for the present model  $ 1/4 < d
< 1/2$ and $2/3 < \eta < 1$.

 We conclude this latter with a brief description of the physical
properties expected for a system described by model (1,2). First, the
magnetic excitations are spin triplets and spin singlets, as at half
filling \cite{ners}.  As for a general  $S = 1$
magnet, the triplet excitations have a gap (the Haldane gap)
which we call $M_t$. There is a singlet gap $M_0$ as well; the corresponding
excitations are spin and charge
singlets and can be treated  as RVB (resonance valence bond)
excitations.  Remaing gapless mode is the charge
free boson which gives a dramatic enchancement to SC
fluctuations or, if the criterion $M_tM_0 > 0$ is not
met, to CDW. We expect a real SC
transition in the  quasi-three-dimensional systems.

\section{Acknowledgements}
We are grateful to A. A. Nersesyan for the valuable discussions and
interest to the work and to J.-S. Caux for the help in preparing the
manuscript.


\begin{thebibliography}{99}
\bibitem{1} Here and thereafter we define
one dimensional SC as a divergence  in the pairing
susceptibility at T = 0.
\end{thebibliography}
\end{document}